\documentclass[sigconf]{acmart}
\AtBeginDocument{%
  \providecommand\BibTeX{{%
    \normalfont B\kern-0.5em{\scshape i\kern-0.25em b}\kern-0.8em\TeX}}}

\copyrightyear{2022}
\acmYear{2022}
\setcopyright{acmcopyright}\acmConference[WebSci '22]{14th ACM Web Science Conference 2022}{June 26--29, 2022}{Barcelona, Spain}
\acmBooktitle{14th ACM Web Science Conference 2022 (WebSci '22), June 26--29, 2022, Barcelona, Spain}
\acmPrice{15.00}
\acmDOI{10.1145/3501247.3531539}
\acmISBN{978-1-4503-9191-7/22/06}

\usepackage{enumitem}
\usepackage{multirow}
\usepackage{graphicx}

\begin{document}

\title{BlackLivesMatter 2020: An Analysis of Deleted and Suspended Users in Twitter}

\author{Cagri Toraman}
\orcid{}
\affiliation{%
  \institution{Aselsan Research Center}
  \city{Ankara}
  \country{Turkey}
}
\email{ctoraman@aselsan.com.tr}

\author{Furkan \c{S}ahinu\c{c}}
\affiliation{%
  \institution{Aselsan Research Center}
  \city{Ankara}
  \country{Turkey}
}
\email{fsahinuc@aselsan.com.tr}

\author{Eyup Halit Yilmaz}
\affiliation{%
  \institution{Aselsan Research Center}
  \city{Ankara}
  \country{Turkey}
}
\email{ehyilmaz@aselsan.com.tr}

\renewcommand{\shortauthors}{Toraman et al.}

\begin{abstract}
After George Floyd's death in May 2020, the volume of discussion in social media increased dramatically. A series of protests followed this tragic event, called as the 2020 BlackLivesMatter movement. Eventually, many user accounts are deleted by their owners or suspended due to violating the rules of social media platforms. In this study, we analyze what happened in Twitter before and after the event triggers with respect to deleted and suspended users. We create a novel dataset that includes approximately 500k users sharing 20m tweets, half of whom actively participated in the 2020 BlackLivesMatter discussion, but some of them were deleted or suspended later. We particularly examine the factors for undesirable behavior in terms of spamming, negative language, hate speech, and misinformation spread. We find that the users who participated to the 2020 BlackLivesMatter discussion have more negative and undesirable tweets, compared to the users who did not. Furthermore, the number of new accounts in Twitter increased significantly after the trigger event occurred, yet new users are more oriented to have undesirable tweets, compared to old ones.
\end{abstract}




\maketitle
\section{Introduction}\label{section:intro}

Social media is a tool for people discussing and sharing opinions, as well as getting information and news. A series of protests in social media followed George Floyd's death in May 2020, called the 2020 BlackLivesMatter (BLM) movement. People participated in the discussion for several reasons. Some users had advocacy to protest racism and police violence. Some tried to take advantage of the popularity of the event by spamming and misinformation spread. Chaos emerged due to many users participating in discussion with different motivations. Eventually, particular accounts are deleted by users or suspended due to violating the rules of social media platforms.

\begin{figure}[t]
\centering
\includegraphics[scale=0.53]{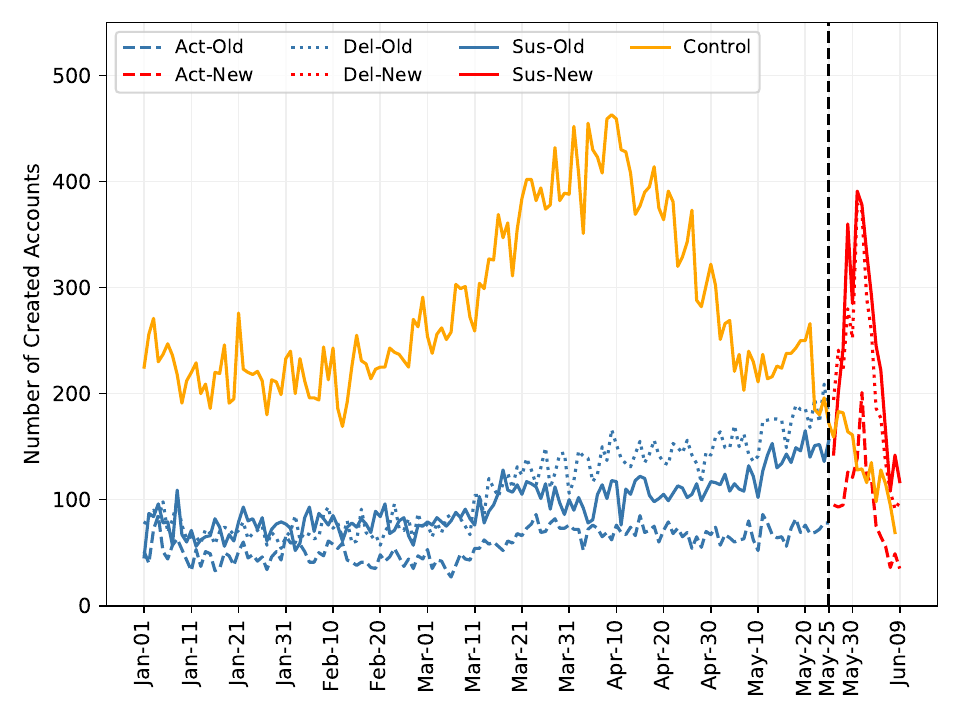} 
\caption{Distribution of the number of new users per day in Twitter (best viewed in color). May 25, 2020 is the day of the event that triggers the 2020 BLM movement. \emph{Act/Del/Sus-Old/New} refers to active, deleted, and suspended users who participated in the BLM discussion and joined Twitter before and after May 25, respectively. Control group is the set of users who do not share any BLM-related tweet.}
\label{fig:acc_dates}
\end{figure}

We analyze the number of new users in Twitter before and after the death of George Floyd in Figure \ref{fig:acc_dates}, using a novel tweet dataset that we describe in the next section. We notice that the number of new users increased dramatically after the event; however most of them are either deleted or suspended later. Meanwhile, there is a strong decrease in control users who do not share any BLM-related tweet. These observations direct us to analyze what happened in social media before and after the event triggers in terms of deleted and suspended users. We thereby examine eight user types; i.e. active, deleted, suspended, and control users who joined the discussion before and after the event (old and new users).
For instance, we refer to suspended users who joined after May 25 as \emph{Sus-New}. Our study has three research questions: 

\textbf{RQ1:} What differences exist between the users who did participate and not participate in the BLM discussion? 

\textbf{RQ2:} What differences exist between the users who joined the discussion before and after the start of the 2020 BLM movement? 

\textbf{RQ3:} What are the possible reasons for user deletion and suspension related to the 2020 BLM movement?

We analyze deleted and suspended users in terms of four factors. 
First, there are social spammers \cite{Lee:2010} who mostly promote irrelevant content, such as abusing hashtags for advertisements and promoting irrelevant URLs. Second, some users have tendency towards deletion possibly due to the regret of sharing negative sentiments \cite{Sleeper:2013}. Third, there are haters and trolls who use offensive language and hate speech mostly against people who have different demographic background \cite{Basile:2019,Silva:2016}. Lastly, social media is an important tool for the circulation of fake news as observed during the 2016 US Election \cite{Fourney:2017}, and for dis/misinformation spread \cite{Wu:2019}; such that Twitter reveals user accounts connected to a propaganda effort by a Russian government-linked organization during the 2016 Election \cite{Twitter:2019}. We refer to misinformation as a term including all types of false information (i.e. dis/misinformation, conspiracy, hoax, etc.). 

The contributions of our study are that we (i) create a novel dataset, called \emph{BlackLivesMatter2020}, that includes over 500k users sharing approximately 20m tweets half of whom actively participated in the 2020 BLM discussion, but some of them were deleted or suspended later; (ii) provide an analysis of the 2020 BLM movement on social media, supported by statistical tests, with a detailed focus on deleted and suspended users who joined the platform before and after the event triggers. 

\section{Related Work}

\subsection{Deletion and Suspension}
Deletion and suspension in social media are analyzed in different cases rather than 2020 BLM. \citet{Thomas:2011} inspect suspended accounts that are responsible for spamming activities such as account selling and ad-based URL shortening. The reason of suspension can be a result of coordinated spam activities during the 2016 and 2020 U.S. election period \cite{Le:2019,Chowdhury:2021}. Similarly, there are studies to understand why users delete contents and accounts \cite{Zhou:2016}.

Supervised prediction models are proposed for deleted \cite{Zhou:2016}, suspended \cite{Volkova:2017,Seyler:2021}, and even troll accounts \cite{Im:2020}. However, users who join social media before and after the start of a social movement are not studied in terms of active, deleted, and suspended user accounts. We fill this gap by providing a comprehensive analysis on a novel dataset regarding 2020 BLM. Our findings can provide a basis for understanding and filtering undesirable behavior in social media after critical events occur.

\subsection{Factors of Undesirable Behavior}
Attacking other people and communities in social media has different forms, such as hate speech and offensive language \cite{Silva:2016, Basile:2019, Davidson:2019}. Deep learning models outperform traditional machine learning with hand-crafted features to detect hate speech text \cite{Cao:2020,Caselli:2021}. In this study, we rely on a recent Transformer language model, RoBERTa \cite{Liu:2019}, which shows state-of-the-art performances for hate speech detection \cite{Zampieri:2020} and sentiment analysis \cite{Barbieri:2020}.

Identification and verification of online information is a recent research area, including fact-checking \cite{Parekh:2020} and fake news detection \cite{Giachanou:2020}. Following similar studies \cite{Chowdhury:2021}, we identify misinformation by using a hand-crafted set of fake and rumor keywords.

Spammers are widely studied in online social networks; including social spammers \cite{Lee:2010} and social bots \cite{Ferrara:2016}. \citet{Chowdhury:2020} emphasize that the malicious behavior of purged users is not limited to spamming. Spreading controversial political content frequently is also a major factor for suspension. In this study, we consider two types of spammers who share similar content multiple times (i.e. social spammers and bots), and promote irrelevant content (i.e. spamming URLs and click-bait).

\section{B\texorpdfstring{\MakeLowercase{lack}}\texorpdfstring{L}\texorpdfstring{\MakeLowercase{ives}}\texorpdfstring{M}\texorpdfstring{\MakeLowercase{atter2020}}\texorpdfstring{:} A Novel Tweet Dataset}\label{section:dataset}

\subsection{Data Preparation}

We created our dataset in two steps. First, we collected all tweets from Twitter API's public stream between April 07 and June 15, 2020. From this collection, we found a user set who posted tweets about the 2020 BLM movement by using a list of 68 BLM-related keywords, such as \#BlackLivesMatter and \#BLM. We manually identify the popular hashtags and keywords shared online during 2020 BLM. We did not omit tweets in languages other than English as long as they contain the keywords, e.g. \#BlackLivesMatter is a global hashtag. The vast majority of the tweets are in English; followed by Spanish, Portuguese, French, and Indonesian. Language distribution displays similar patterns for different user types. We also collect tweets for control users who do not share any BLM-related keywords during sampling period.

In December 2020, we checked the status of the user accounts. Some of the accounts maintain their activity. Here, the term ``activity" is used to indicate the existence of an account. Users may post tweets actively, or remain as listeners. There are also deleted accounts, and the remaining ones are suspended by Twitter. In the second step, we expand the tweets by fetching the archived tweets of all users from Wayback Machine\footnote{https://archive.org/details/twitterstream}. To obtain a better insight on user activity before and after the start of the 2020 BLM movement, we collected the archived tweets posted two months before and after the event.

\begin{table}[t]
\fontsize{9}{10.8}\selectfont
\caption{Distribution of users and tweets by the user types, along with the average number of tweets per user, and special elements per tweet (hashtags and URLs).}
\label{tab:user_tweet_counts}
\centering
\begin{tabular}{|l|r|r|r|c|c|}

\hline
\multicolumn{1}{|c|}{\multirow{2}{*}{\textbf{Type}}} & \multicolumn{1}{|c|}{\multirow{2}{*}{\textbf{User}}} & \multicolumn{1}{|c|}{\multirow{2}{*}{\textbf{Tweet}}} & \multicolumn{3}{c|}{\textbf{Average}} \\\cline{4-6}
& & & \textbf{Tweet} & \textbf{Htag} & \textbf{URL} \\
\hline
Cont-Old & 247,475 & 9,578,561 & 38.71 & 0.29 & 0.05 \\ \hline

Cont-New & 2,517 & 77,989 & 35.36 & 0.43 & 0.03 \\ \hline

Act-Old & 118,363 & 4,035,769 & 34.10 & 0.28 & 0.11 \\ \hline

Act-New & 1,637 & 22,870 & 13.97 & 0.33 & 0.10 \\ \hline

Del-Old & 66,775 & 2,319,065 & 34.73 & 0.21 & 0.07 \\ \hline

Del-New & 3,982 & 48,592 & 12.20 & 0.33 & 0.10 \\ \hline

Sus-Old & 54,839 & 3,696,509 & 67.41 & 0.29 & 0.10 \\ \hline

Sus-New & 4,646 & 82,035 & 17.66 & 0.35 & 0.11 \\ \hline

TOTAL & 500,234 & 19,861,390 & 39.70 & 0.28 & 0.07 \\ \hline
\end{tabular}
\end{table}

\subsection{Descriptive Statistics}
Our dataset\footnote{The dataset includes publicly available user and tweet IDs, in compliance with Twitter's Terms and Conditions, and can be accessed, along with our BLM-related keywords, from https://github.com/avaapm/BlackLivesMatter} includes 500,234 users with 19,861,390 tweets; 250,242 users participated in the BLM discussion at least once, while the remaining users did not participate in the BLM discussion at all (i.e. the control group). We acknowledge that deleted and suspended users and their tweets can not be retrieved from Twitter API, however we share user and tweet IDs to provide transparency and reliability to our study.

Novelty of the dataset comes from both its content and size. To the best of out knowledge, there is no other BLM related tweet dataset at this size including labels of control, active, deleted, and suspended user labels. 

We give the distribution of users and tweets in Table \ref{tab:user_tweet_counts}, along with the average number of tweets per user, and the average number of hashtags and URLs per tweet. To provide a fair analysis, we collect half of the tweets for control, whereas the other half of the tweets are distributed equally as much as possible among active, deleted, and suspended users. The average number of tweets per user is 39.70 in overall. The average number of hashtags shared by new accounts is higher than those of old accounts. The new accounts of deleted and suspended share more URLs compared to the old ones.

\section{Experiments}
\label{section:experiments}

In this section, we first explain the design of our experiments, then report the experimental results.

\subsection{Experimental Design}
In order to answer the RQs, we design our experiments in terms of hashtags and URLs, spamming, sentiment analysis, hate speech, and misinformation spread. We use multilingual tweets in all experiments, except that we filter non-English tweets in sentiment analysis and hate speech by using the language field provided by Twitter API.

\subsubsection{Hashtags and URLs}
We report the most frequent hashtags and URL domains shared in the tweets to observe any abnormal activity. We fact-check the most frequent URL domains in terms of false information by using PolitiFact\footnote{https://www.politifact.com}. We check the scorecard of the domain in PolitiFact and if the number of fact checks for ``mostly false", ``false" and ``pants on fire" is higher than number of ``true" checks, then we assign domain as suspicious one.

\subsubsection{Spamming} 
Spam behavior is mostly observed when users share duplicate and similar content multiple times \cite{Hinesley:2019}, and exploit popular hashtags to promote their irrelevant content, i.e. hashtag hijacking \cite{VanDam:2016}. We detect both issues and find spam tweets. To detect the former, we find near-duplicate tweets by measuring higher than 95\% text similarity between tweets using the Cosine similarity with TF-IDF term weighting. To detect the latter, we group popular hashtags into topics, and label a tweet as spam if it contains hashtags from different sets. The topics and related keywords are obtained from \cite{Sahinuc:2021}. The topics are the BLM movement, COVID-19 pandemic, Korean Music, Bollywood, Games\&Tech, and U.S. Politics. Since the number of tweets is large, we can not label the ground truth. However, we manually check the results and observe no significant amount of false positives in our preliminary experiments.

\subsubsection{Sentiment Analysis}
Users can express their opinion in either a negative or positive way. We apply RoBERTa \cite{Liu:2019} fine-tuned for the task of sentiment analysis in terms of polarity classification (positive-negative-neutral) \cite{Barbieri:2020} to the tweet contents.
We filter non-English tweets, and tweets with less than three tokens excluding hashtags, links, and user mentions. Any existing duplicate tweets due to retweets are removed as well. Preprocessing results in keeping approximately \%60 of tweets in BLM-related users, and \%18 for non-BLM. 
    
\subsubsection{Hate Speech}
During the discussion of important events, some users can behave aggressively and even use hate speech towards other individuals or groups of people. We apply RoBERTa \cite{Liu:2019} fine-tuned for the task of detecting hate speech and offensive language \cite{Mathew:2020} to the tweet contents. The same text filtering and cleaning steps used in the sentiment analysis experiments are also followed in this experiment.
    
\subsubsection{Misinformation Spread}
Misinformation for the 2020 BLM movement can be examined under a number of topics. Since hashtag is one of important instruments of misinformation campaigns \cite{Arif:2018}, we find tweets with potential misinformation containing a list of hashtags and keywords regarding five misinformation topics. We observe no significant amount of false positives in our preliminary experiments. Topics are as follows:  (i) Authorities block protestors to communicate by blackout in Washington DC, (ii) A black teenager is violently arrested by US police, (iii) George Soros funded the protests, (iv) ANTIFA organized the protests, and (v) St. Paul police officer Jacob Pederson is the provocateur who helped to start the looting.

For each user type, $t{\epsilon}T$, where $T=\{\emph{Act,Del,Sus,Cont}\}{\times}\{\emph{Old,New}\}$, we split the tweet set into k independent subsets. In each subset, we measure \emph{factor ratio}, the average number of tweets belonging to a factor, as follows.

\begin{equation}
\text{f}_t = (n_{t,f})/(n_t/k)
\end{equation}

\noindent where $f\epsilon\{\emph{spam,negative,hate,misinfo}\}$, $n_{t,f}$ is the number of tweets assigned to the factor $f$ in the user type $t$, and $n_t$ is the number of all tweets in the user type $t$. We set \emph{k} to 100 for the experiments. We determine statistically significant differences between the mean of 100-observation subsets ($\mu_{t,f}$) following non-normal distributions by using the two-sided Mann-Whitney U (MWU) test at \%95 interval with Bonferroni correction. To scale the results in the plots, we report \emph{normalized factor ratio} for the user type $t$ as follows.

\begin{equation}
 \text{nf}_t = (\mu_{t,f})/(\sum_{i\epsilon{T}}{\mu_{i,f}})
\end{equation}

\begin{figure*}[ht]
\centering
\includegraphics[width=1.6\columnwidth]{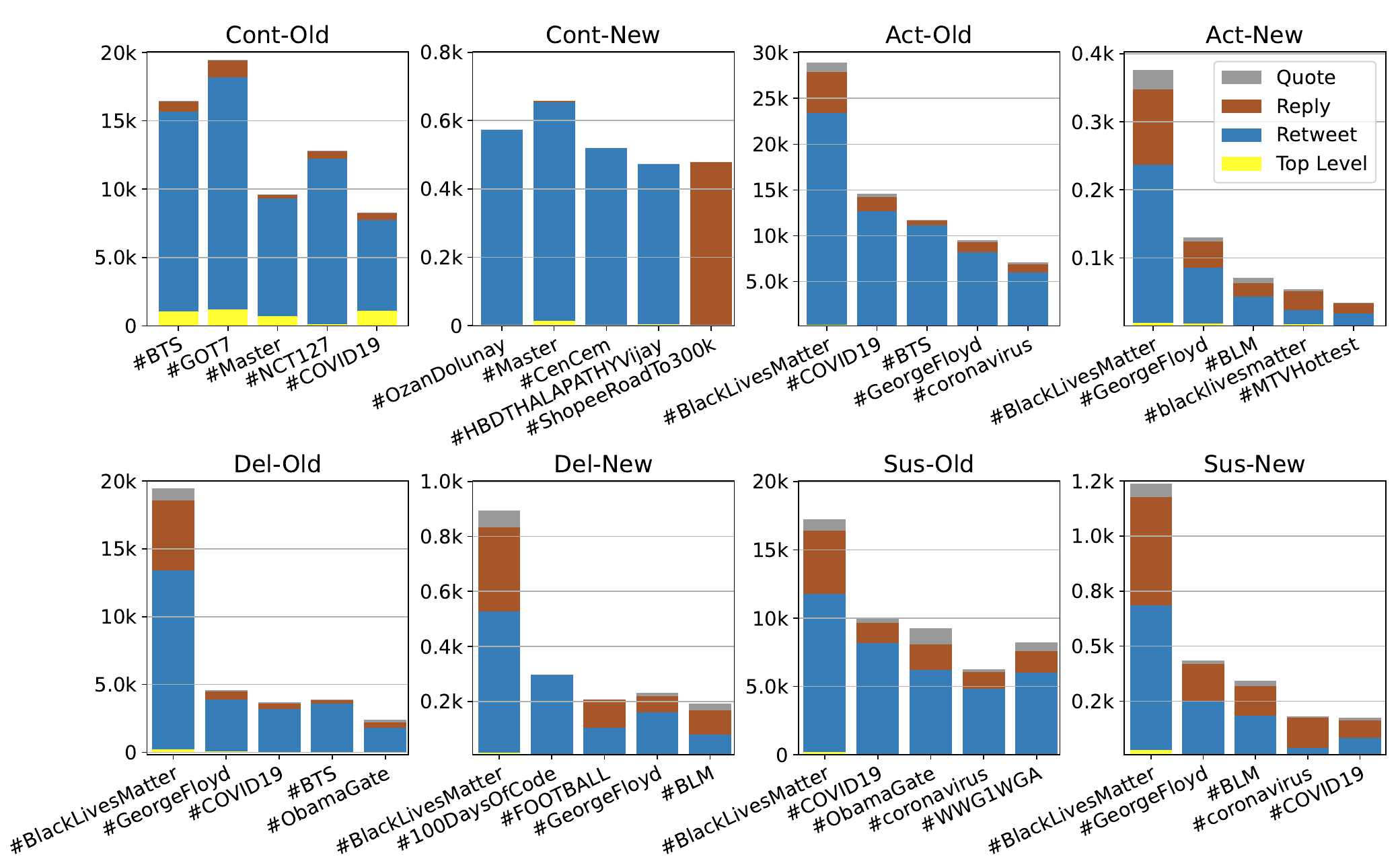}
\caption{Frequent hashtags shared in the tweets for each user type, ranked according to hashtag frequency (best viewed in color). Multiple occurrences in a tweet are ignored in ranking but not in individual bars.} 
\label{fig:htags}
\end{figure*}

\subsection{Experimental Results}
In this section, we report our experimental results to answer the RQs. 

\begin{table*}[ht]
\fontsize{9}{10.8}\selectfont
\caption{Frequent URL domains in the tweets for each user type, ranked according to frequency ratio. Domain extensions (.com and .co.uk) are removed for simplicity. Suspicious ones that could spread false info are given in bold. Note that suspicious domains do not always spread misinformation. We decide suspicious ones by using the scorecard values in PolitiFact.}
\label{tab:url_statistics}
\centering
\setlength{\tabcolsep}{1pt}
\begin{tabular}{|l|l|l|l|l|l|l|l|l|l|l|l|l|l|l|l|}
\hline
\multicolumn{2}{|c}{\textbf{Cont-Old}} &
\multicolumn{2}{|c}{\textbf{Cont-New}} &
\multicolumn{2}{|c}{\textbf{Act-Old}} & \multicolumn{2}{|c}{\textbf{Act-New}} & \multicolumn{2}{|c}{\textbf{Del-Old}} & \multicolumn{2}{|c}{\textbf{Del-New}} & \multicolumn{2}{|c}{\textbf{Sus-Old}} & \multicolumn{2}{|c|}{\textbf{Sus-New}} \\
\hline
\cline{1-16} \cline{1-16} \cline{1-16} \cline{1-16} 
.12 & youtube & .11 & youtube & .07 & youtube & .09 & youtube & .09 & youtube & .08 & youtube & .09 & youtube & .11 & youtube \\

.02 & onlyfans & .08 & blogspot & .02 & nytimes & .07 & itechnews & .02 & change.org & .03 & thepugilistmag & .02 & \textbf{thegatewaypundit} & .05 & allthelyrics \\

.02 & instagram & .04 & instagram & .02 & washingtonpost & .03 & wesupportpm & .02 & spotify & .02 & carrd.co & .02 & \textbf{foxnews} & .02 & cloudwaysapps \\

.02 & spotify & .04 & onlyfans & .01 & cnn & .02 & cnn & .02 & \textbf{foxnews} & .02 & change.org & .02 & \textbf{breitbart} & .02 & \textbf{foxnews} \\

.02 & peing.net & .03 & spotify & .01 & instagram & .02 & change.org & .02 & onlyfans & .02 & openionsblog & .01 & nytimes & .01 & \textbf{breitbart} \\

\hline
\end{tabular}
\end{table*}

\subsubsection{Hashtags and URLs}
We report the most frequent hashtags shared in the tweets in Figure \ref{fig:htags}. \emph{\#BlackLivesMatter} is the most frequent hashtag for each user type, except control users. Since the data is collected by using a set of keywords, the highest numbers of observed hashtags in all user types except the control group belong to the BLM topic. Nevertheless, we observe hashtags that belong to different topics as well. Control users (non-BLM) share hashtags from various topics with no single dominant hashtag (RQ1). Old users share not only BLM-related hashtags, but also other popular topics during the same time period, such as \emph{\#COVID19} and \emph{\#ObamaGate}, while new users share mostly BLM-related hashtags (RQ2). New users actively participate in the discussion, since the ratio of hashtags in replies is higher, compared to old users, as observed in the barchart. 

For old and suspended users, we observe more right-wing political hashtags, such as \emph{\#WWG1WGA} and \emph{\#ObamaGate}. Suspended users could be more oriented to misinformation campaigns, compared to active and deleted users (RQ3). The ratio of hashtags in quotes is higher for them compared to other hashtags. 

We report the most frequent URL domains in Table \ref{tab:url_statistics}.
Similar to hashtags, we observe mostly right-wing political domains for suspended users. A right-wing political domain, \emph{foxnews.com}, is shared more by the deleted and suspended users, compared to the active users who share \emph{nytimes.com}, \emph{washingtonpost.com}, and \emph{cnn.com}. In deleted users, \emph{change.org} is consistently observed.

\begin{figure*}[t]
\centering
\includegraphics[width=1.7\columnwidth]{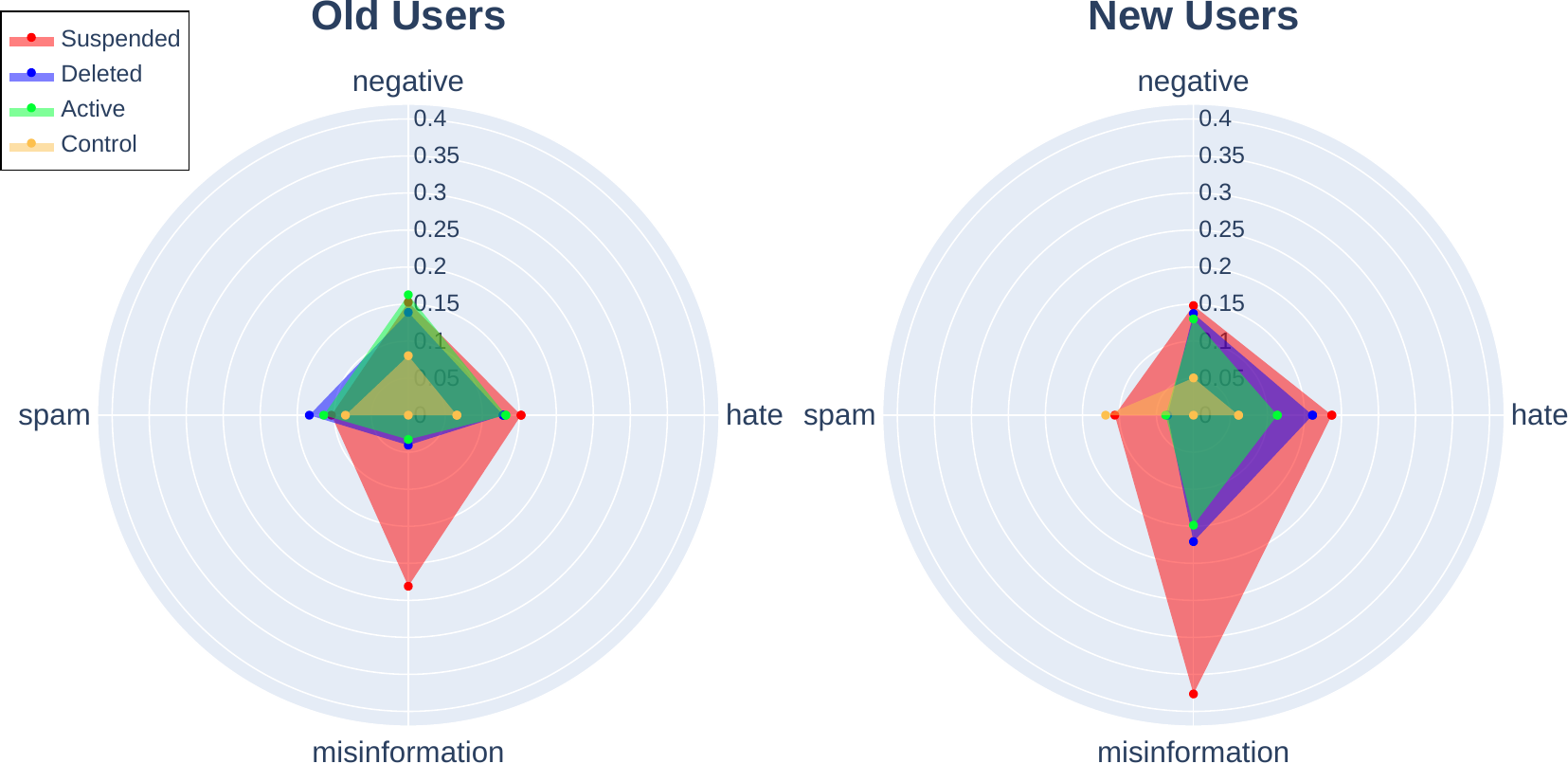} 
\caption{Normalized factor ratio for spamming, negative sentiment, hate speech, and misinformation spread of old and new users (best viewed in color). Misinformation is not available for Control that has non-BLM users.}
\label{fig:radar_subplots}
\end{figure*}

\subsubsection{Factors for Deletion and Suspension}
We report normalized factor ratio in terms of spamming, negative sentiment, hate speech, and misinformation in Figure \ref{fig:radar_subplots}. Note that misinformation scores are not available for Control group, since the topics are about BLM and Control group has no BLM-related tweets. 

When BLM and non-BLM users are compared (RQ1); BLM users have more spam, more negative, and more hate speech tweets than non-BLM (control) users (MWU-test, Bonferroni adjusted p$<$0.008); except the spamming behavior of new users. 

When old and new users are compared (RQ2), new users have more negative, more hate speech, and more tweets on misinformation topics, compared to old users. These differences are statistically significant (MWU-test, Bonferroni adjusted p$<$0.0125); except deleted users for hate, and misinformation. This observation supports that, new users (specifically suspended ones) are more oriented to use negative sentiments, hate speech, and misinformation during important social movements, compared to old users. Yet, there is no statistically significant change regardless of users being old and new in hate speech, and misinformation behavior of deleted users; and spamming behavior of suspended ones.

When user types are compared (RQ3), suspended accounts share more spamming, more negative, more hate speech, and more tweets on misinformation topics, compared to active and deleted users, specifically for new users (MWU-test, Bonferroni adjusted p$<$0.008). Considering old accounts, suspended users have more hate speech and misinformation tweets, but not negative and spam tweets. We observe many negative tweets of active users, which might support that such users yield to protest the event. Another observation is that deleted users have more negative and hate speech, compared to active users, specifically for new accounts (MWU-test, Bonferroni adjusted p$<$0.008).

\section{Discussion}
\label{section:discussion}

\subsection{Main Insights}
From our experimental results, we argue that global events, such as the 2020 BlackLivesMatter (BLM) movement, can have a significant impact on people's behaviours in social media, such that users involve in the discussion more. On the other hand, this momentum in social media creates an opportunity for other users who display undesirable behaviors, such as promoting misinformation and spamming irrelevant content. 

Users who participated in the 2020 BLM discussion have more negative tweets, and more undesirable behavior (i.e. spamming, hate speech, and misinformation spread), compared to the users who did not participate (RQ1).
    
The number of new accounts in Twitter increased significantly after the event that triggers 2020 BLM. New users mostly participated in the BLM discussion, and they are more oriented to have undesirable behavior, compared to old users (RQ2). We emphasize the presence of new users for hate speech and misinformation detection. Deleted users are an exception for this observation. 
    
Suspended users have more undesirable tweets, compared to active and deleted users (RQ3). This observation is consistent with the rules of the social media platforms. Although the platforms attack undesirable behavior extensively, we still find such tweets that are shared by active users, specifically new ones. We argue that regulations on new account creation could be further discussed, as well as monitoring new accounts during important social events.

\subsection{Limitations}
We acknowledge some limitations to our study. The data is collected by filtering a hand-crafted set of keywords, resulting in missing some related tweets. The data for the control group is down-sampled to have a closer number of tweets to those of BLM-related tweets, causing potential information loss. Moreover, keyword matching to distinguish control users from BLM-related ones can cause the existence of some noisy instances, which can be further removed by dedicated noise reduction methods in future. Hate speech detection relies on the performance of RoBERTa \cite{Liu:2019}, and misinformation detection on the performance of keyword matching. We have possibly some bias in the results due to incorrectly classified and missed tweets. Since the data is too large, we can not label the ground truth and evaluate accordingly, however we observe no significant amount of such cases in the preliminary experiments.

\subsection{Ethical Considerations}
We state that the data we work on is collected from the sources that are publicly available. There is no data collection from the private accounts. In addition, collected data are not used to obtain any kind of additional information from other sources. In general, the standards from \citet{Rivers:2014} are followed. In order to provide transparency, we publish online\footnote{https://github.com/avaapm/BlackLivesMatter} the details of dataset (e.g. language distribution per user type), experimental design (e.g. the keywords used for misinformation spread), and experimental results (e.g. the list of normalized factor ratio values and the details of statistical tests).

\section{Conclusion}
We analyze what happened in Twitter before and after the event that triggers the 2020 BLM movement on a novel dataset with approximately 500k users and 20m tweets including deleted and suspended users, called \emph{BlackLivesMatter2020}. We report substantial differences between old and new users participated in 2020 BLM, and reasons for user deletion and suspension. The experimental results are supported by statistical tests. Our analysis is based on a novel tweet dataset; yet the results can be extended to other social media platforms and different case studies. 

The main observations that we report in the 2020 BLM event can be compared with similar events to understand the generalization of this analysis. The features analyzed in this study can be further exploited in predicting user deletion and suspension. Bot accounts can be analyzed in details using our dataset in the future.

\bibliographystyle{ACM-Reference-Format}
\bibliography{bibliography}

\end{document}